\documentclass{aa}

\usepackage{graphicx}
\usepackage[varg]{txfonts}
\usepackage{natbib}
\bibpunct{(}{)}{;}{a}{}{,}
\usepackage{array}
\usepackage{xspace}
\usepackage{lscape}
\usepackage{url}
\usepackage{arydshln}
\usepackage[hyperfootnotes=false, linktocpage=true, breaklinks=true, colorlinks=true, linkcolor=blue, citecolor=blue, urlcolor=blue]{hyperref}
\usepackage[all]{hypcap}
\usepackage{longtable}
\usepackage{afterpage}
\usepackage{multirow}

\usepackage{enumitem}
\newlist{myenumerate}{enumerate}{1}
\setlist[myenumerate]{label=$^{\arabic*}$}

\newcommand{\fcov}{{\ensuremath{C_f}}\xspace}

\newcommand{\FeKb}{Fe K\ensuremath{\beta}\xspace}
\newcommand{\kms}{\ensuremath{\mathrm{km\ s^{-1}}}\xspace}
\newcommand{\NH}{\ensuremath{N_{\mathrm{H}}}\xspace}

\newcommand{\xabs}{\xspace{\tt xabs}\xspace}
\newcommand{\xmm}{{\it XMM-Newton}\xspace}
\newcommand{\chandra}{{\it Chandra}\xspace}
\newcommand{\swift}{{\it Swift}\xspace}

\newcommand{\hst}{{\it HST}\xspace}

\newcommand{\ngc}{{NGC~3783}\xspace}

\newcommand{\nustar}{{\it NuSTAR}\xspace}

\newcommand{\ergcm}{{\ensuremath{\rm{erg\ cm}^{-2}\ \rm{s}^{-1}\ {\AA}^{-1}}}\xspace}
\newcommand{\ergflux}{{\ensuremath{\rm{erg\ cm}^{-2}\ \rm{s}^{-1}}}\xspace}
\newcommand{\cm}{{\ensuremath{\rm{cm}^{-2}}}\xspace}

\newcommand{\spex}{\xspace{\tt SPEX}\xspace}
\newcommand{\pion}{\xspace{\tt pion}\xspace}

\newcommand{\logxi}{\ensuremath{{\log \xi}}\xspace}

\mathchardef\mhyphen="2D

\begin{document}

\title{Chasing obscuration in type-I AGN: discovery of an eclipsing clumpy wind at the outer broad-line region of NGC 3783}

\author{
M. Mehdipour \inst{1}
\and
J.S. Kaastra \inst{1,2}
\and
G.A. Kriss \inst{3}
\and
N. Arav \inst{4}
\and
E. Behar \inst{5}
\and 
S. Bianchi \inst{6}
\and
G. Branduardi-Raymont \inst{7}
\and
M. Cappi \inst{8}
\and
E. Costantini \inst{1}
\and
J. Ebrero \inst{9}
\and
L. Di Gesu \inst{10}
\and
S. Kaspi \inst{5}
\and
J. Mao \inst{1,2}
\and
B. De Marco \inst{11}
\and
G. Matt \inst{6}
\and
S. Paltani \inst{10}
\and
U. Peretz \inst{5}
\and
B.M. Peterson \inst{3,12,13}
\and 
P.-O. Petrucci \inst{14,15}
\and
C. Pinto \inst{16}
\and
G. Ponti \inst{17}
\and
F. Ursini \inst{8}
\and
C.P. de Vries \inst{1}
\and
D.J. Walton \inst{16}
}

\institute{
SRON Netherlands Institute for Space Research, Sorbonnelaan 2, 3584 CA Utrecht, the Netherlands\\ \email{M.Mehdipour@sron.nl}
\and
Leiden Observatory, Leiden University, PO Box 9513, 2300 RA Leiden, the Netherlands
\and
Space Telescope Science Institute, 3700 San Martin Drive, Baltimore, MD 21218, USA
\and
Department of Physics, Virginia Tech, Blacksburg, VA 24061, USA
\and
Department of Physics, Technion-Israel Institute of Technology, 32000 Haifa, Israel
\and
Dipartimento di Matematica e Fisica, Universit\`{a} degli Studi Roma Tre, via della Vasca Navale 84, 00146 Roma, Italy
\and
Mullard Space Science Laboratory, University College London, Holmbury St. Mary, Dorking, Surrey, RH5 6NT, UK
\and
INAF-IASF Bologna, Via Gobetti 101, I-40129 Bologna, Italy
\and
European Space Astronomy Centre, P.O. Box 78, E-28691 Villanueva de la Ca\~{n}ada, Madrid, Spain
\and
Department of Astronomy, University of Geneva, 16 Ch. d'Ecogia, 1290 Versoix, Switzerland
\and
Nicolaus Copernicus Astronomical Center, Polish Academy of Sciences, Bartycka 18, PL-00-716 Warsaw, Poland
\and
Department of Astronomy, The Ohio State University, 140 West 18th Avenue, Columbus, OH 43210, USA
\and
Center for Cosmology \& AstroParticle Physics, The Ohio State University, 191 West Woodruff Ave., Columbus, OH 43210, USA
\and
Univ. Grenoble Alpes, IPAG, F-38000 Grenoble, France
\and
CNRS, IPAG, F-38000 Grenoble, France
\and
Institute of Astronomy, Madingley Road, CB3 0HA Cambridge, UK
\and
Max Planck Institute fur Extraterrestriche Physik, 85748, Garching, Germany
}

\date{Received 15 May 2017 / Accepted 12 July 2017}

\abstract
{
In 2016 we carried out a \swift monitoring program to track the X-ray hardness variability of eight type-I AGN over a year. The purpose of this monitoring was to find intense obscuration events in AGN, and thereby study them by triggering joint \xmm, \nustar, and \hst observations. We successfully accomplished this for NGC~3783 in December 2016. We found heavy X-ray absorption produced by an obscuring outflow in this AGN. As a result of this obscuration, interesting absorption features appear in the UV and X-ray spectra, which are not present in the previous epochs. Namely, the obscuration produces broad and blue-shifted UV absorption lines of Ly$\alpha$, \ion{C}{iv}, and \ion{N}{v}, together with a new high-ionisation component producing \ion{Fe}{xxv} and \ion{Fe}{xxvi} absorption lines. In soft X-rays, only narrow emission lines stand out above the diminished continuum as they are not absorbed by the obscurer. Our analysis shows that the obscurer partially covers the central source with a column density of few $10^{23}$~cm$^{-2}$, outflowing with a velocity of few thousand km~s$^{-1}$. The obscuration in \ngc is variable and lasts for about a month. Unlike the commonly-seen warm-absorber winds at pc-scale distances from the black hole, the eclipsing wind in \ngc is located at about 10 light days. Our results suggest the obscuration is produced by an inhomogeneous and clumpy medium, consistent with clouds in the base of a radiatively-driven disk wind at the outer broad-line region of the AGN.
}
\authorrunning{M. Mehdipour et al.}
\titlerunning{Chasing obscuration in type-I AGN: discovery in NGC 3783}
\maketitle

\section{Introduction}
Accretion onto supermassive black holes (SMBHs) in active galactic nuclei (AGN) is believed to be accompanied by outflows of gas, which couple the SMBHs to their environment. The observed associations between SMBHs and their host galaxies, such as the {M-$\sigma$ relation} \citep{Ferr00}, point to their co-evolution through a feedback mechanism. The AGN outflows may play an important role in this feedback as they can impact star formation, chemical enrichment of the intergalactic medium, and cooling flows in galaxy clusters (e.g. review by \citealt{Fabi12}). There are however significant gaps in our understanding of the outflow phenomenon in AGN.

Winds of photoionised gas ({warm absorbers} - WA) are commonly observed in bright AGN through high-resolution UV and X-ray spectroscopy (e.g. \citealt{Cren99, Blu05}). They often consist of multiple ionisation components, outflowing with velocities of typically few hundred \kms. From an observational point of view, other kinds of winds with different properties from WAs have been found in the X-ray band: {high-ionisation ultra-fast outflows} (e.g. \object{PDS~456}, \citealt{Reev09}) and {obscuring outflows} (e.g. \object{NGC~5548}, \citealt{Kaas14}). Compared to the common WAs at pc-scale distances from the black hole (e.g. \citealt{Kaas12}), the obscuring outflow found in NGC~5548 is a faster and more massive wind closer to the accretion disk. It produces strong absorption of the X-ray continuum, in addition to appearance of blue-shifted and broad UV absorption lines. X-ray obscuration with associated UV line absorption has been seen also in \object{Mrk~335} \citep{Long13} and \object{NGC~985} \citep{Ebre16}. Variable X-ray absorption is commonly found in type-I AGN; e.g. \object{NGC~1365} \citep{Rive15}; \object{PDS~456} \citep{Matz16}; \object{NGC~4151} \citep{Beuc17}; \object{IRAS~13224-3809} \citep{Park17}. However, the association to the UV broad-line absorbing outflows is unclear. Moreover, the physical connection between different kinds of AGN outflow, and their origin and driving mechanism, are still poorly understood. In this study we aim to address the nature and origin of an X-ray obscuration/eclipse through UV/X-ray spectroscopy of the absorption during an eclipsing event.

An efficient way to drive winds in quasars is via radiative acceleration of the gas through UV line absorption (e.g. \citealt{Prog04}). However, intense X-ray radiation from the central source can over-ionise the gas, leaving insufficient line opacity to drive the wind. Shielding the UV-absorbing gas from the X-rays by an obscuring medium near the X-ray source (like that seen in NGC~5548) can prevent this. Thus, obscuration may play an important role in driving AGN outflows. A statistical study of X-ray variability by \citet{Mark14} identifies obscuration events in AGN using RXTE observations. They find 12 X-ray eclipses in 8 AGN, and compute the probability of finding a type-I AGN undergoing obscuration is $\sim$1\%. However, the origin, location, and physical properties of such eclipses are poorly understood. It is also uncertain whether such eclipses are a manifestation of disk winds in general. In order to broaden our understanding of this phenomenon, we have conducted a \swift monitoring program on a sample of type-I AGN to catch an obscuration event and perform a ToO multiwavelength spectroscopic study of it using \xmm, \nustar, and \hst COS.

\section{\textit{\textbf{Swift}} monitoring program and triggering of \textit{\textbf{XMM-Newton, NuSTAR, and HST}} observations}
\label{data_sect}

The X-ray spectral hardness variability is a useful indicator of obscuration. We define the hardness ratio (HR) as ${(H-S) / (H+S)}$, where $H$ and $S$ are the \swift XRT count rates in the hard (1.5--10 keV) and soft (0.3--1.5 keV) bands, respectively. X-ray absorption by obscuring/eclipsing gas increases HR. During \swift Cycle 12 (April 2016--March 2017), we monitored eight suitable type I AGN: \object{Ark~564}, \object{MR~2251-178}, Mrk~335, \object{Mrk~509}, \object{Mrk~841}, NGC~3783, \object{NGC~4593}, and \object{NGC~7469}. These AGN were observed weekly by \swift during the corresponding visibility windows of the four observatories. While most of the AGN displayed stable HR throughout the year, only \ngc (triggered by us) and Mrk~335 (triggered earlier by another team) showed significant X-ray spectral hardening.

\begin{figure}[!tbp]
\centering
\hspace*{-0.6cm}\resizebox{1.125\hsize}{!}{\includegraphics[angle=0]{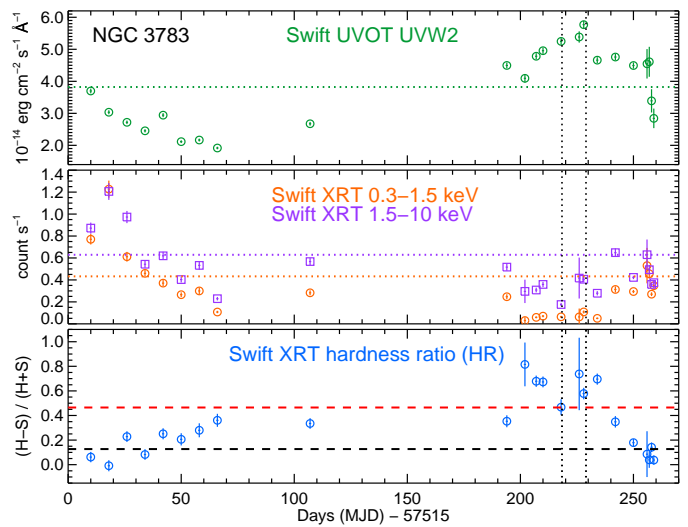}}
\caption{\swift lightcurve of \ngc from 17 May 2016 to 21 January 2017. The horizontal dotted lines in the two upper panels show the all-time average \swift flux levels. The dashed black line in the bottom panel indicates the average quiescent hardness ratio (HR) from unobscured data. The dashed line in red is the HR limit for triggering, above which significant obscuration was predicted according to our simulations. The first and second \xmm observations are indicated by vertical dotted lines.}
\label{swift_lc}
\end{figure}

\begin{figure}[!tbp]
\centering
\hspace*{-0.35cm}\resizebox{1.05\hsize}{!}{\includegraphics[angle=0]{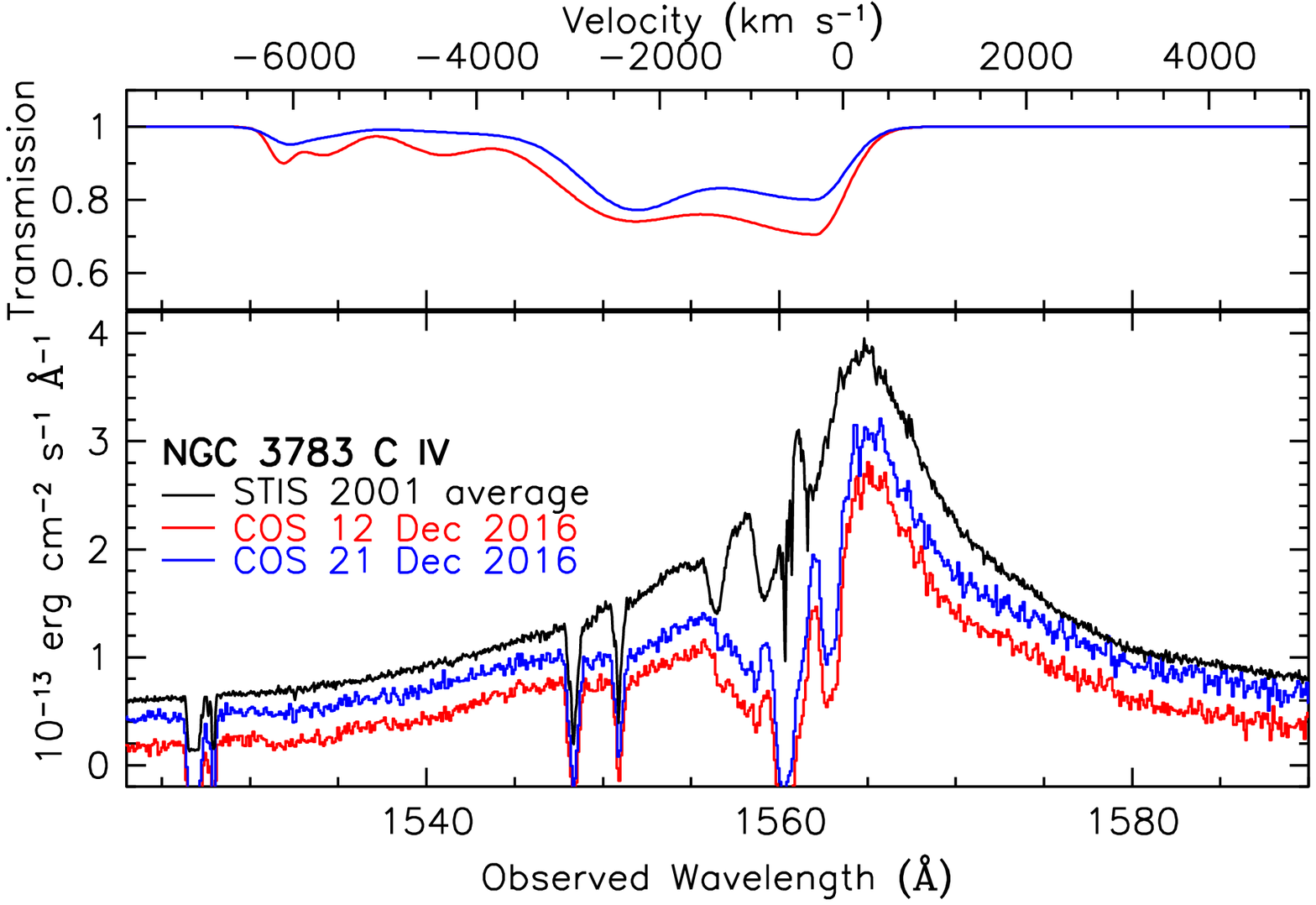}}\vspace{-6.7cm}
\caption{\ngc~\hst COS and STIS spectra near the \ion{C}{iv} line. The line transmission model for the new broad \ion{C}{iv} absorption component in 2016 is shown in the top panel. The displayed spectra are continuum subtracted and then offset vertically by $2.5 \times 10^{-14}$~\ergcm for each epoch so that the weaker changes in the absorption are more visible. Narrow absorption lines in the blue wing are interstellar \ion{Si}{ii}~$\lambda1526$ and \ion{C}{iv}~$\lambda1548,1551$.}
\label{cos_fig}
\end{figure}

Figure \ref{swift_lc} shows the \swift lightcurve of NGC 3783 from May 2016 to January 2017. In December 2016, we found an intense X-ray spectral hardening event lasting for about 32 days. During this period we successfully executed the triggering of our \xmm, \nustar, and \hst observations (see Table \ref{obs_table} in Appendix \ref{appendix}). Figure \ref{spectra_fig} (upper panel) shows the 2016 \xmm EPIC-pn and \nustar spectra, as well as the time-averaged EPIC-pn spectra from 2000 and 2001. Strong X-ray absorption is evident in the new data (see also the RGS data in Fig. \ref{spectra_fig}, bottom panel), with the 0.3--2.0 keV flux dropping from ${1.60 \times 10^{-11}}$~\ergflux in 2000--2001 by a factor of 8.0 (11 Dec 2016) and 4.5 (21 Dec 2016). This X-ray absorption coincides with an increase in the UV flux (Fig.~\ref{swift_lc}). Strong line absorption affects the blue side of the \ion{C}{iv} line profile in the 2016 \hst/COS spectrum (Fig. \ref{cos_fig}), extending from the line center to ${\sim -3200}$~\kms, with additional shallow absorption features present down to ${\sim -6200}$~\kms. Blue-shifted broad UV line absorption is also detected in Ly$\alpha$ and \ion{N}{v} in the new COS spectra \citep{Kris17}.

For a description of our data reduction, we refer to Appendix A in \citet{Meh15a}, which applies to the \ngc data used here, with more details provided in our follow-up papers. The wavelength/energy bands used in our simultaneous X-ray spectral modelling of the data are 7--37~$\AA$ for RGS, 1.5--10 keV for EPIC-pn, and 10--80 keV for \nustar. The spectral modelling is done using the {\tt SPEX} package v3.03.01 \citep{Kaa96}. We use C-statistics for spectral fitting with X-ray spectra optimally binned according to \citet{Kaas16}. Errors are reported at $1\sigma$ confidence level.

%

%
\begin{figure}[!tbp]
\vspace{-0.43cm}
\centering
\hspace*{-0.78cm}\resizebox{1.21\hsize}{!}{\includegraphics[angle=270]{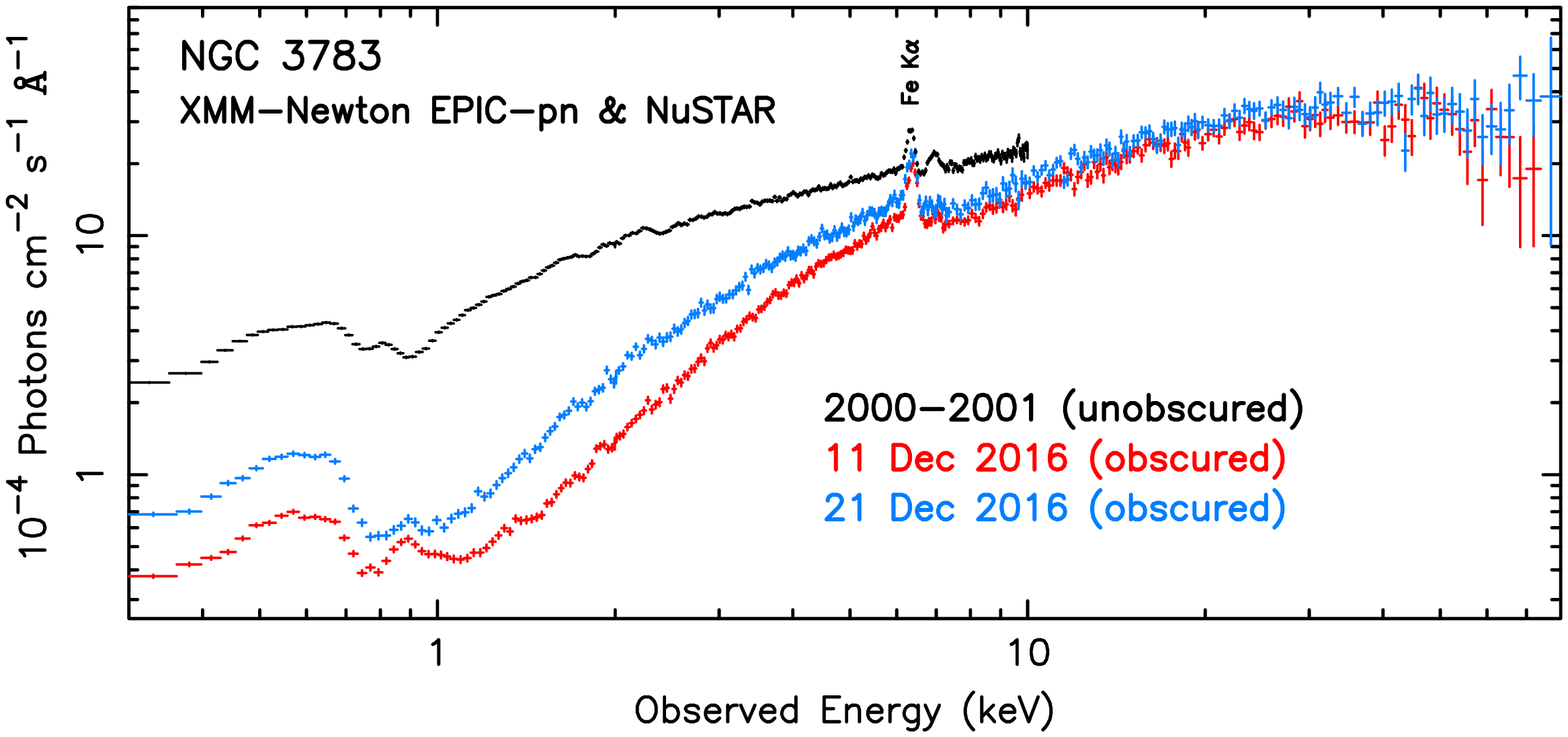}}\vspace{-3.5cm}
\hspace*{-0.78cm}\resizebox{1.21\hsize}{!}{\includegraphics[angle=270]{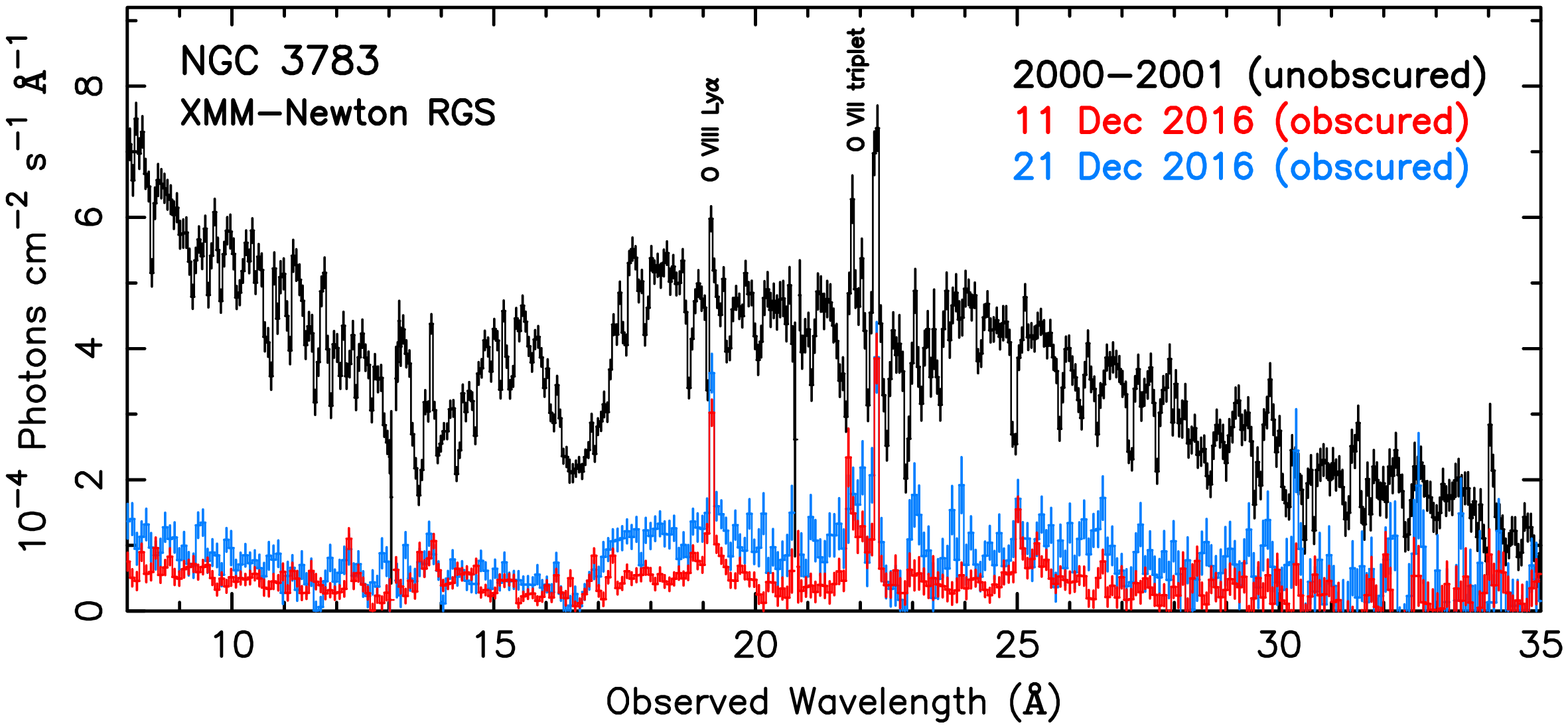}}\vspace{-2.8cm}
\caption{\ngc spectra from \xmm EPIC-pn and \nustar ({\it top panel}), and \xmm RGS ({\it bottom panel}). The displayed energy range for EPIC-pn is 0.3--10 keV and for \nustar 10--80 keV.
}
\label{spectra_fig}
\vspace{-0.3cm}
\end{figure}

%

\section{Modelling of the obscuring wind in \ngc}
\label{SED_sect}

For photoionisation modelling of the WA and the new obscurer, we determined the spectral energy distribution (SED) of the central ionising source in \ngc. We applied a template SED model that we reported in \citet{Meh15a} for NGC~5548 to fit the \ngc data and determine its SED. These Seyfert-1 AGN have a SED composition consisting of an optical/UV thin disk component, an X-ray power-law continuum, a neutral X-ray reflection component, and a warm Comptonisation component for the soft X-ray excess. The exponential cut-off energy of the power-law was set to 340~keV \citep{DeRo02}, which is also consistent with the \nustar spectra. The Galactic X-ray absorption is modelled using the {\tt hot} model in \spex, with $\NH = 9.59 \times 10^{20}$~cm$^{-2}$ \citep{Murp96}. The redshift of \ngc is set to 0.009730 \citep{Theu98}, and all abundances are fixed to the proto-solar values of \citet{Lod09}. To correct for Galactic reddening, we used the {\tt ebv} model, with ${E(B - V) = 0.107}$ \citep{Schl11}. To take into account the host galaxy optical/UV stellar emission, we used the galactic bulge model of \citet{Kin96}, and normalised it to the \ngc host galaxy flux measured from \hst \citep{Ben13}. In the $12\arcsec$ diameter circular aperture of OM, this is ${7.04 \times 10^{-15}}$ \ergcm (Bentz, priv. comm.).

Before modelling the new strong absorption by the obscurer in the 2016 data, we first derived a model for the WA from archival observations, where the WA absorption features are clearly detectable in X-rays. Previous studies have found a WA in \ngc \citep{Kasp02, Blu02, Beh03, Scot14}. We utilised all archival data from \xmm (2000 and 2001) and \chandra HETGS (2000, 2001, and 2013) to produce a set of time-averaged spectra. The HETGS spectra were obtained from {\tt TGCat} \citep{Huen11}. For photoionisation and spectral modelling of the optically-thin WA, we used the new {\tt pion} model in \spex \citep{Meh16b}. From modelling the \ngc archival spectrum, we find that the WA spans a wide range of ionisation, similar to the distribution reported by \citet{Holc07} and \citet{Goos16}. We fit the absorption by the WA with multiple {\tt pion} components, with outflow velocities ranging from 450 to 1200~\kms. The narrow X-ray emission lines are also fitted with the \pion model at zero net velocity. The total \NH of the WA is derived to be about ${4.0 \times 10^{22}}$~\cm. More details about this WA model will be reported by \citet{Mao17}.

The 2016 data suggest that the photoionised emission from the X-ray narrow line region is not absorbed by the obscurer. This is evident from the clear presence of narrow emission lines and radiative recombination edges in the RGS spectrum, such as the \ion{O}{viii} Ly$\alpha$ at 19~$\AA$ and \ion{O}{vii} triplet lines at 22~$\AA$ (Fig. \ref{spectra_fig}, bottom panel). We make a reasonable assumption that the obscurer in \ngc is likely located interior to both the WA and the X-ray narrow line region. Previous studies find the WA in \ngc to be at pc-scale distances from the black hole (e.g. \citealt{Beh03, Gabe05}). From our photoionisation modelling, we find that the ionisation state and turbulent velocity of both the WA and the X-ray narrow line region match each other. These indicate that they are likely at similar distances from the black hole, albeit there are modelling uncertainties associated with this interpretation. In our line of sight, the WA is effectively shielded from receiving some of the ionising radiation, thus it becomes less ionised. This lower ionisation is directly evidenced by the increased absorption in the narrow UV outflow components in the 0 to $-1500$~\kms range of the COS spectra in Fig. \ref{cos_fig}. For the WA of \ngc with an electron density of ${3 \times 10^{4}}$~cm$^{-3}$ \citep{Gabe05}, we find the recombination timescale for relevant ions is ${\lesssim 1}$ day. This means during the month-long obscuration event, the WA would be able to respond to the change in the ionising SED caused by the obscuration. Thus, we take into account the enhanced absorption by this {de-ionised} WA. The WA model obtained from the unobscured data is incorporated in our modelling of the new obscured data, with only the ionisation parameter $\xi$ \citep{Kro81} of the WA components self-consistently lowered by the obscuration.

Continuum absorption by the obscurer is too strong to leave detectable absorption lines in soft X-rays. Therefore, to set the velocity and $\xi$ of the obscurer in our modelling, we use the broad UV absorption lines of the obscurer seen in the COS spectrum. The transmission model for the broad \ion{C}{iv} line \citep{Kris17} is shown in Fig. \ref{cos_fig}, top panel. We use the weighted average velocity of the broad \ion{C}{iV} absorption profile in our X-ray spectral modelling (${v_{\rm out} = -1900}$~\kms and ${\sigma_{v} = 1100}$~\kms). The ionisation balance of the obscurer is derived using the {\tt Cloudy} v13.04 photoionisation code \citep{Fer13} for an optically-thick medium, to match the UV lines in the COS spectrum, with its X-ray absorption fitted using the {\tt xabs} model in \spex. This is in order to produce the observed UV lines without a massive neutral hydrogen front as there are no significant detections of \ion{C}{ii} and \ion{Si}{ii} in the COS spectrum of \ngc. We obtain a solution at ${\log \xi =1.84}$ for the obscurer.

%
\begin{figure}[!tbp]
\centering
\hspace*{-0.4cm}\resizebox{0.98\hsize}{!}{\includegraphics[angle=0]{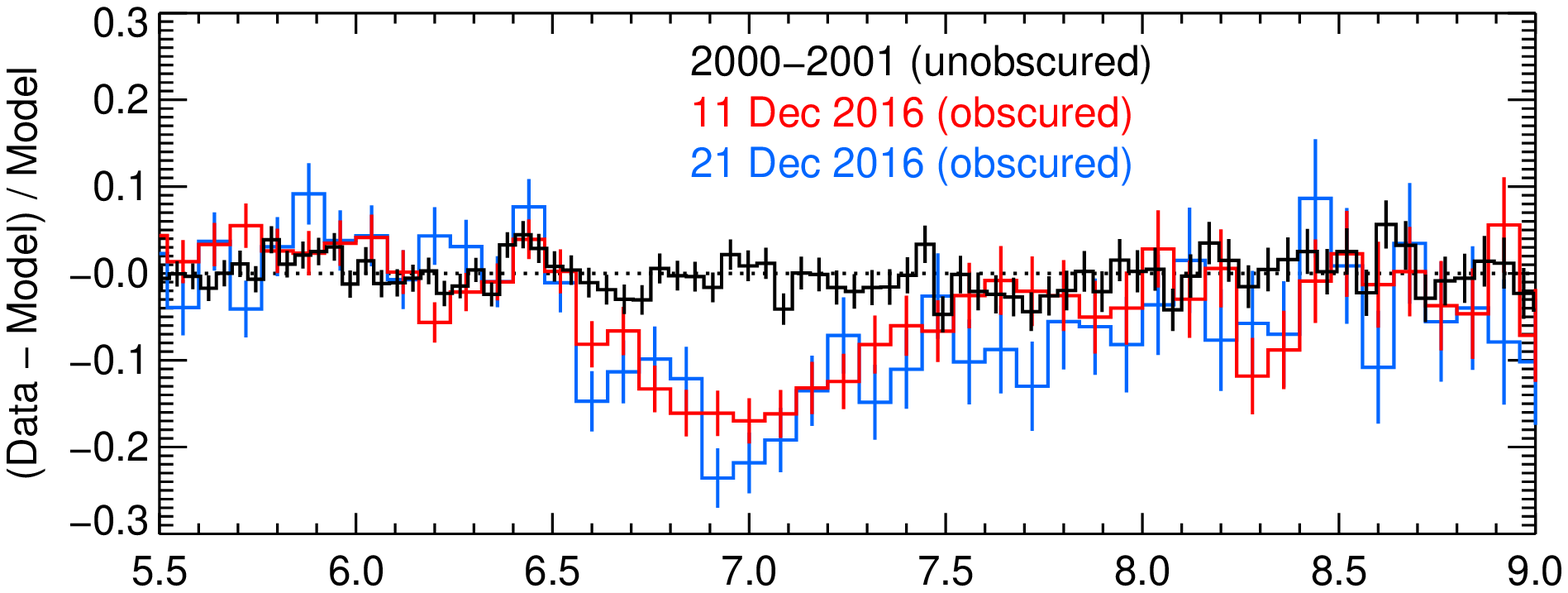}}\vspace{-0.7cm}
\hspace*{-0.4cm}\resizebox{0.98\hsize}{!}{\includegraphics[angle=0]{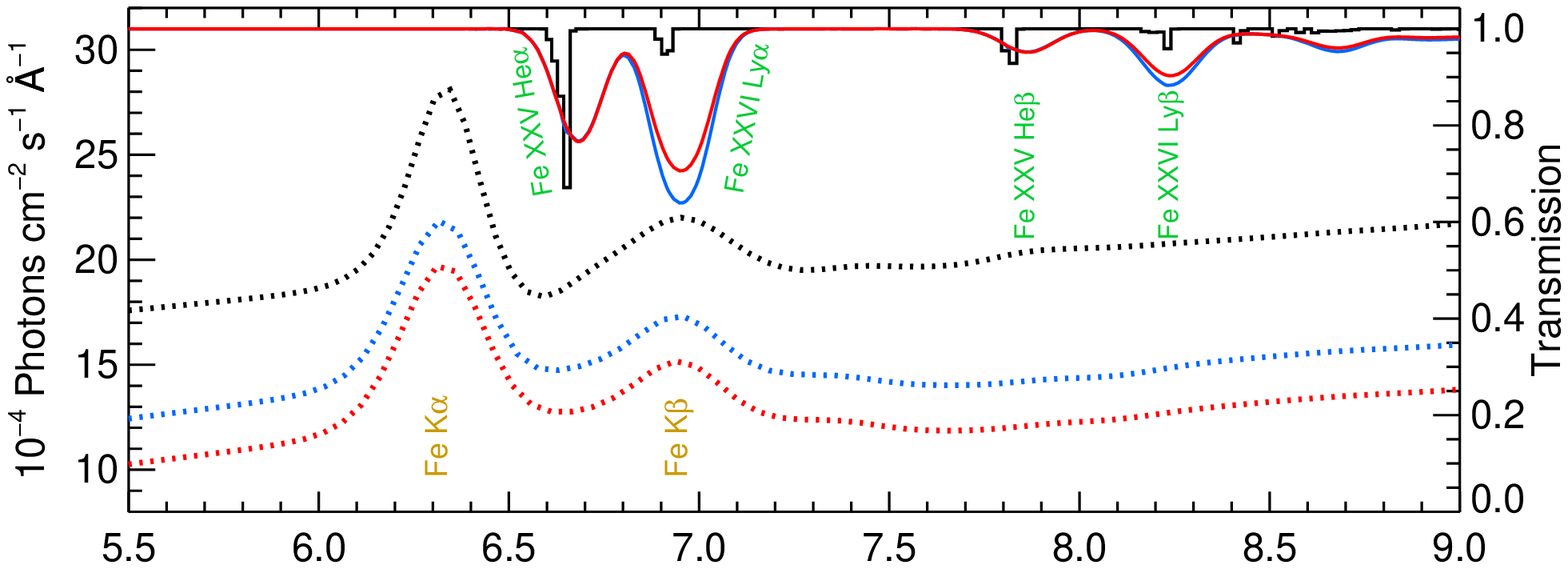}}\vspace{-0.7cm}
\hspace*{-0.4cm}\resizebox{0.98\hsize}{!}{\includegraphics[angle=0]{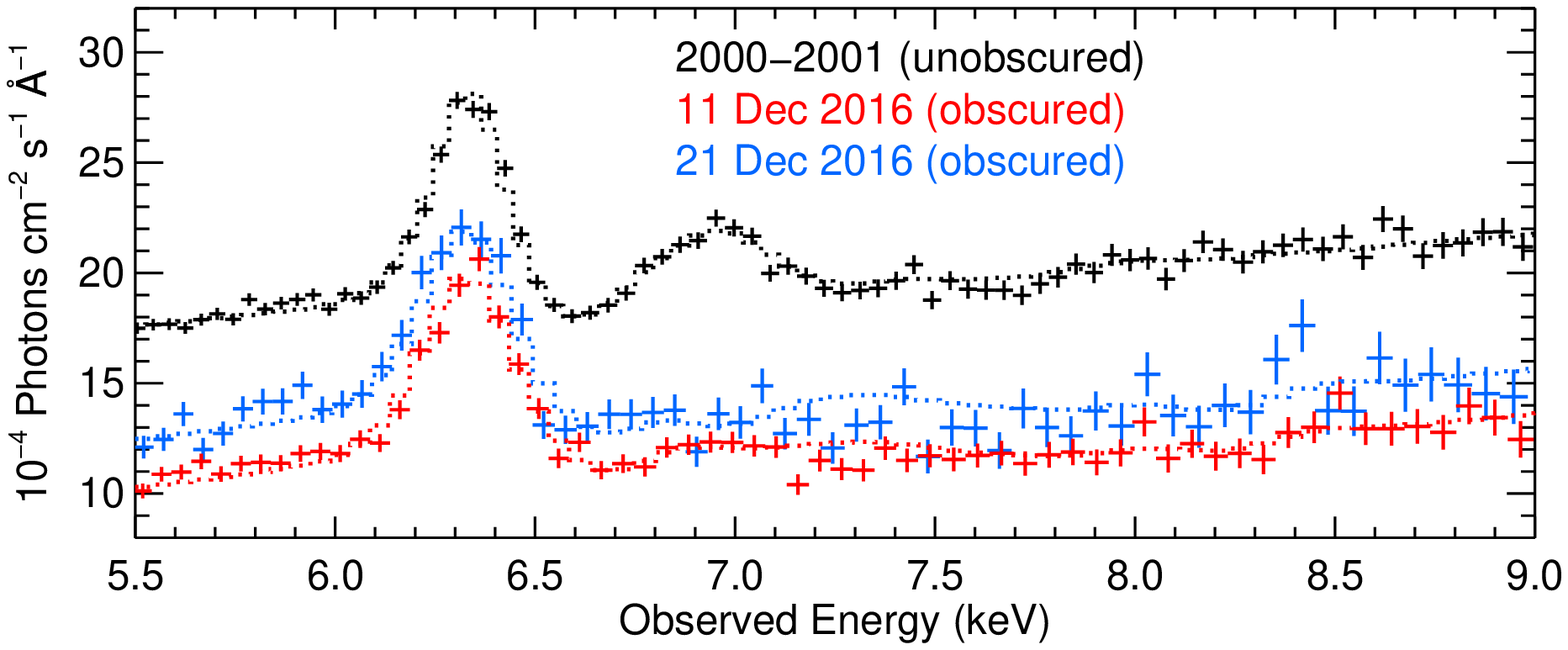}}
\caption{{\it Top panel:} \ngc EPIC-pn data fit residuals without absorption by a high-ionisation component (HC), showing new \ion{Fe}{xxv} and \ion{Fe}{xxvi} absorption lines in the 2016 obscured epoch. {\it Middle panel}: Line transmission of the blue-shifted HC \ion{Fe}{xxv} and \ion{Fe}{xxvi} absorption lines (solid lines), shown with respect to the Fe K$\alpha$ and Fe K$\beta$ emission lines and the continuum unaffected by the HC line absorption (dotted lines). {\it Bottom panel}: EPIC-pn data and their best-fit model (dotted lines).}
\label{residual_fig}
\vspace{-0.2cm}
\end{figure}

%
\begin{figure}[!tbp]
\centering
\hspace*{-0.35cm}\resizebox{1.06\hsize}{!}{\includegraphics[angle=0]{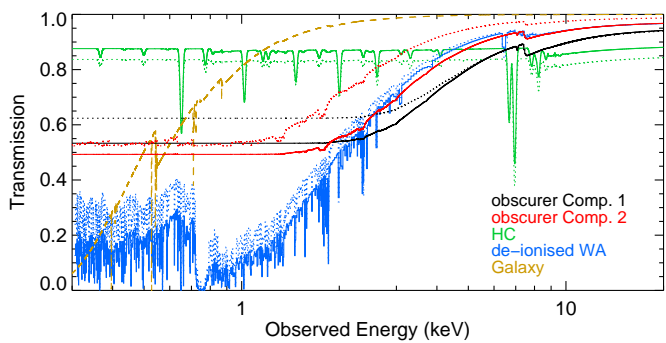}}
\caption{Continuum and line transmission of all the absorption components in our line of sight towards the nucleus of \ngc. The solid lines correspond to Obs. 1, and the dotted lines to Obs. 2.}
\label{trans_fig}
\end{figure}

\begin{figure}[!tbp]
\centering
\hspace*{-0.17cm}\resizebox{1.035\hsize}{!}{\includegraphics[angle=0]{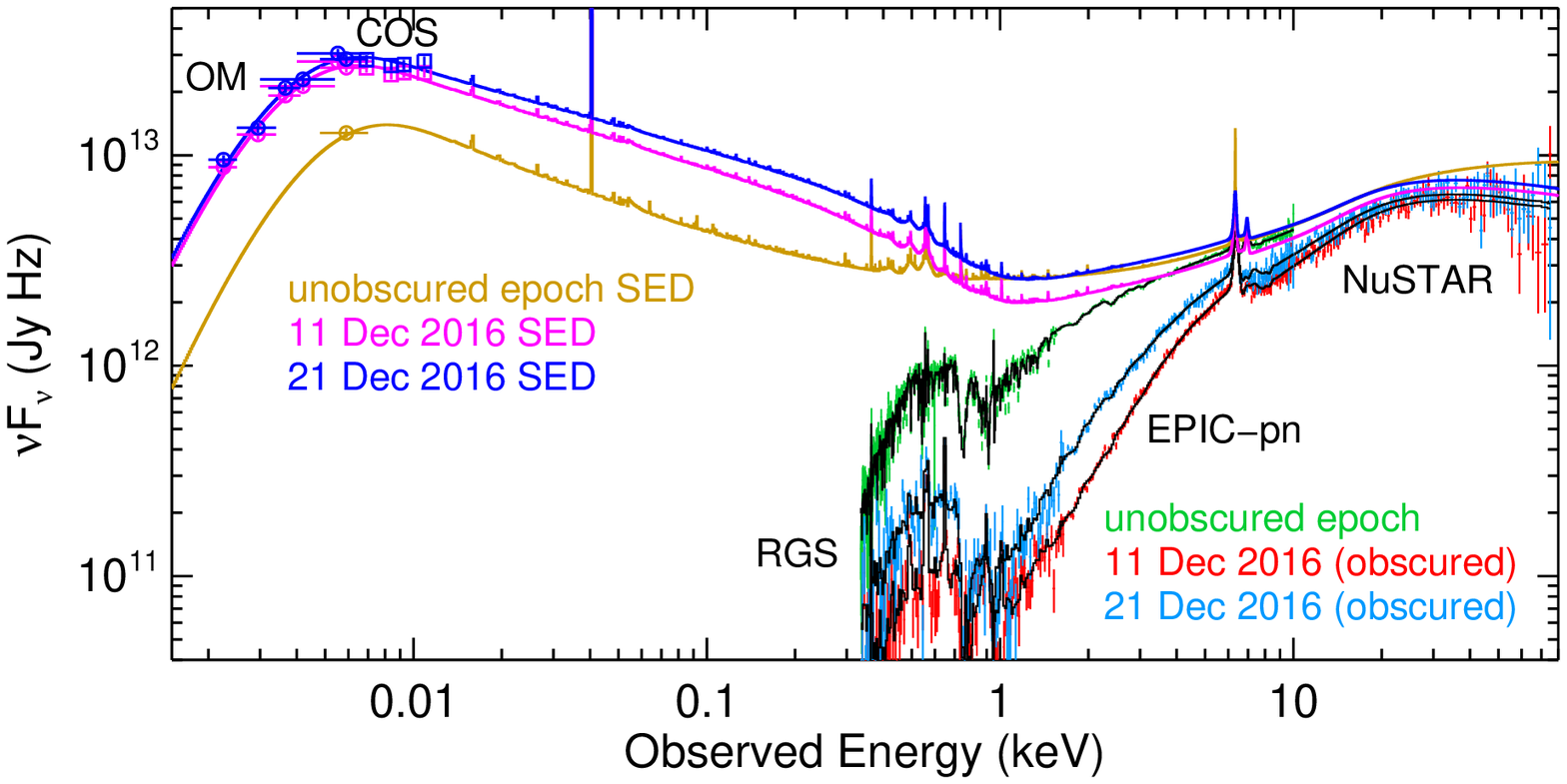}}\vspace{-0.4cm}
\hspace*{-0.17cm}\resizebox{1.035\hsize}{!}{\includegraphics[angle=0]{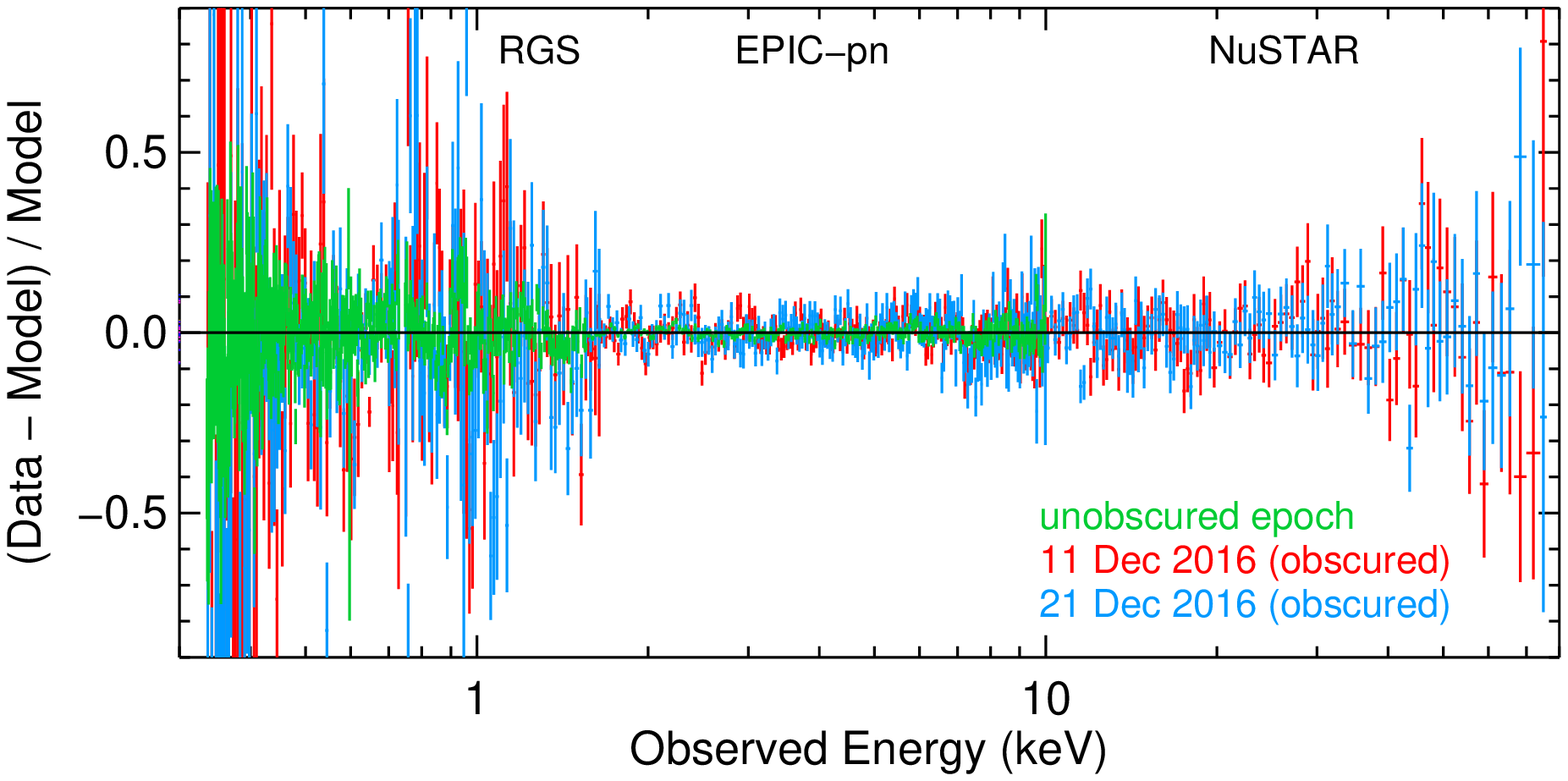}}
\caption{{\it Top panel}: Best-fit model to each dataset of \ngc, and its corresponding SED model. The displayed X-ray data include absorption by the Galaxy, the WA, and the obscurer. {\it Bottom panel}: Residuals of the best-fit model.}
\label{sed_fig}
\end{figure}

To fit the 2016 obscured X-ray spectra we require two {\xabs} absorption components to reproduce the observed curvature of the spectrum from soft to hard X-rays. The curvature seen in 2016 is not present in the archival spectra. The addition of the first and second {\xabs} components improve the fit significantly with $\Delta {\rm C}$ of about 4000 and 1000, respectively. Interestingly, we find evidence for a strong high-ionisation component (HC) in the 2016 data (see Fig. \ref{residual_fig}). The \ion{Fe}{xxvi}~Ly$\alpha$ line (${E_{0} = 6.966}$~keV), blue-shifted by about $-2300$~\kms, overlaps with the \FeKb emission line (${E_{0} = 7.020}$~keV). This \ion{Fe}{xxvi} absorption of the continuum causes the \FeKb emission line to vanish in 2016, while it was present in the archival data (Fig. \ref{residual_fig}). The addition of this HC component further improves the fit with $\Delta {\rm C}$ of about 200. The X-ray transmission of all the absorption components in our line of sight to \ngc are shown in Fig. \ref{trans_fig}. The final model fits the data well with C-stat\,/\,d.o.f. = 2288\,/\,1539 (Obs. 1) and 2285\,/\,1542 (Obs. 2). We note that the remaining fit residuals primarily belong to soft X-ray emission lines from the X-ray narrow line region. The modelling of these lines is independent of the obscurer. The obscurer itself is effectively featureless in X-rays (Fig.~\ref{trans_fig}), detected only through continuum absorption and the characteristic curvature in the broadband X-ray continuum. The continuum is fitted well as there are no curvature residuals in our best-fit model (Fig \ref{sed_fig}, bottom panel). The intrinsic photon index $\Gamma$ of the underlying X-ray power-law continuum is found to be about 1.71 (Obs. 1) and 1.75 (Obs. 2). Our best-fit model to the data, and the corresponding SEDs, are displayed in Fig. \ref{sed_fig}. The best-fit parameters of the obscurer and the HC are given in Table \ref{para_table}.

%

\section{Discussion and conclusions}
\label{obscurer_sect}

The strongly diminished soft X-rays and the appearance of new spectral features in \ngc can be explained by an obscuring wind at the core of this AGN. The obscurer is found to consist of two partially-covering absorption components, suggesting the obscuring medium is inhomogeneous and clumpy. A similar property was also found for the obscuring wind in NGC~5548 \citep{Kaas14}. However, in the case of NGC~5548, the obscuration has been lasting for several years \citep{Meh16a}, while the one in \ngc was a short-lived eclipsing event, lasting for about a month (Fig. \ref{swift_lc}). Moreover, the obscurer in \ngc has a significantly higher ionisation parameter $\xi$ and lower covering fraction $C_f$ than the one in NGC~5548. These differences can be explained if the obscurer in NGC~5548 is a spatially extended stream of cool gas in our line of sight \citep{Kaas14}, while the one in \ngc is a hotter transient cloud, which is localised closer to the central source.

%
\begin{table}[!tbp]
\begin{minipage}[t]{\hsize}
\setlength{\extrarowheight}{2pt}
\caption{Best-fit parameters of the two obscurer components and the new high-ionisation component (HC) for Obs. 1 (11 Dec 2016) and Obs. 2 (21 Dec 2016).}
\label{para_table}
\centering
\small
\renewcommand{\footnoterule}{}
\begin{tabular}{c | c | c c c c c c}
\hline \hline
Parameter 						& Obs.	&	Comp. 1				&	Comp. 2	 		& 	HC		        \\
\hline
\multirow{1}{*}{\NH}	 				& 1		&	$1.5\pm0.2$			&	$0.8\pm0.2$		& 	$1.6\pm0.3$	\\
($10^{23}$~cm$^{-2}$)				& 2		&	$2.0\pm0.2$			&	$0.3\pm0.1$		& 	$2.2\pm0.5$		\\
\hline
\multirow{2}{*}{\fcov}					& 1		&	$0.47\pm0.10$			&	$0.51\pm0.10$		& 	$1$ (f) 			\\
				 				& 2		&	$0.38\pm0.03$			&	$0.48\pm0.03$		& 	$1$ (f)			\\
\hline
\multirow{1}{*}{\logxi}					& 1		&	$1.84$ (f)				&	$1.84$ (f)			& 	$3.61\pm0.05$		        \\
(erg~cm~s$^{-1}$) 					& 2		&	$1.84$ (f)				&	$1.84$ (f)			& 	$3.77\pm0.07$		        \\							
\hline
\multirow{1}{*}{$v_{\rm out}$~(\kms)} 	& 1, 2	&	$-1900$ (f)			&	$-1900$ (f)		& 	$-2300$ (f)		\\

\hline
\multirow{1}{*}{$\sigma_{v}$~(\kms)} 	& 1, 2	&	$1100$ (f)				&	$1100$ (f)			& 	$2500$ (f)		\\
												
\hline
\multicolumn{8}{c}{C-stat\,/\,d.o.f. = 2288\,/\,1539 (Obs. 1) and 2285\,/\,1542 (Obs. 2)} \\
\hline
\end{tabular}
\end{minipage}
\end{table}

\ngc has previously displayed X-ray spectral hardening in the archival \swift and RXTE data. There is a single obscured \swift observation in 2009, and from RXTE \citet{Mark14} identify four eclipsing events. However, in 2016 we got the first opportunity to carry out a spectroscopic study of the obscuration with \xmm, \hst, and \nustar. Since the obscurer is found to partially cover the central X-ray source, we may assume it has a transverse size ($d$) comparable to that of the X-ray source. The transverse velocity ($v_{\rm t}$) required by the obscurer to eclipse the X-ray source is $2 R_{\rm X} / t$, where $R_{\rm X}$ is the radius of the X-ray corona and $t$ is the duration of the obscuration event (i.e. 32 days). We adopt a fiducial radius of ${10\,R_{\rm g}}$ for the X-ray corona, where the gravitational radius ${R_{\rm g} = GM_{\bullet}/c^{2}}$, with $G$ the gravitational constant, $c$ the speed of light, and the black hole mass ${M_{\bullet} = 2.98 \times 10^{7}}$~$M_{\odot}$ \citep{Vest06}. These yield ${d  = 8.8 \times 10^{13}}$~cm and ${v_{\rm t} = 320}$~\kms for the obscurer. The obscurer density ${n_{\rm H} \sim \NH /D}$, where $\NH=2.3 \times 10^{23}$~\cm (Table \ref{para_table}) and $D$ is the length of the obscurer in our line of sight. The length $D$ is equal to the above transverse size $d$ assuming an obscuring cloud with a spherical geometry. This gives ${n_{\rm H}}$ of about ${2.6 \times 10^{9}}$~cm$^{-3}$, which is a typical broad-line region (BLR) density (e.g. \citealt{Bald95}). Finally, from the definition of the ionisation parameter $\xi$, we have ${r = \sqrt{L / \xi\, n_{\rm H}}}$, where the ionising luminosity $L$ over 1--1000~Ryd is about ${1.1 \times 10^{44}}$~erg~s$^{-1}$ and ${\xi = 10^{1.84}}$~erg~cm~s$^{-1}$ from our modelling. This yields a distance ${r \sim 10}$ light days from the black hole. For comparison, the radius of the BLR in \ngc from reverberation mapping \citep{Pet04} ranges from about 1.4 (\ion{He}{ii}) to 10.2 (H$\beta$) light days. The radius of the IR torus is 250--357 light days \citep{Beck08}. Therefore, the obscurer is likely located in the outer BLR.

Previous studies of \ngc have found an outflowing HC through the detection of a narrow \ion{Fe}{xxv} He$\alpha$ absorption line (e.g. \citealt{Yaqo05}). From our joint analysis of the stacked archival HETGS and EPIC-pn data, we find the component responsible for this narrow \ion{Fe}{xxv} absorption line has ${\log \xi = 3.0\pm0.1}$ and ${\NH = 1.4\pm0.1 \times 10^{22}}$~\cm, with ${v = -450\pm50}$~\kms and ${\sigma_{v} = 100}$~\kms. However, in the 2016 obscured data a more massive ($2.3 \times 10^{23}$~\cm) and more ionised HC is present, which is also faster and broader than the archival HC (Table~\ref{para_table}). Absorption by the \ion{Fe}{xxvi} Ly$\alpha$ in the HC causes the disappearance of the \FeKb emission line (Fig~\ref{residual_fig}, middle panel). In both archival and obscured observations, the line energy and flux of the Fe~K$\alpha$ line remain unchanged within errors (about $\pm20$~eV in line energy and $\pm10$\% in flux). Therefore, according to the theoretical Fe~K line calculations \citep{Palm03b,Kall04}, similar \FeKb emission line would be expected in 2016, hence its disappearance cannot be due to a change in the ionisation state of the line-emitting region. The appearance of the new HC in 2016 data of \ngc is likely associated with the obscurer. The increase in $\xi$ between the two 2016 observations (Table \ref{para_table}) matches the observed change in the ionising luminosity $L$ of the source between the two observations, varying from 1.0 to ${{1.2 \times 10^{44}}}$~erg~s$^{-1}$. This enables us to put limits on distance $r$ of the HC from its recombination timescale $t_{\rm rec}$. From our photoionisation modelling, $n_{\rm H} \times t_{\rm rec}$ for \ion{Fe}{xxvi} is derived. Since $t_{\rm rec}$ has to be shorter than the spacing between the two observations (9.7 days), this can be used to put constraints on $n_{\rm H}$ and hence $r$. We find the HC has ${n_{\rm H} > 2.3 \times 10^{5}}$~cm$^{-3}$ and ${r < 120}$ light days. Thus, it may coexist spatially with the obscurer at the outer BLR. Both the obscurer and the HC have comparable velocities (Table \ref{para_table}).

The eclipsing obscurer in \ngc is outflowing because it produces blue-shifted and broad absorption lines. Since the obscurer is found to be inhomogeneous and clumpy, and its location matches the BLR, it is consistent with being in the base of a radiatively-driven wind at the BLR \citep{Murr95, Prog04, Prog14}. Similar X-ray eclipses found in NGC~1365 and \object{Mrk~766} have also been attributed to passage of BLR clouds in our line of sight to the X-ray source \citep{Risa07,Risa11}. Similar to \ngc, an association between obscuration and a HC is also found for the luminous quasar PDS~456 \citep{Reev09, Nard15}, where a partially-covering Compton-thick absorber appears together with a highly-ionised relativistic disk wind. Interestingly, the obscuring wind and the HC in this quasar are a more massive and faster version of the wind in the less luminous Seyfert-1 \ngc. We note that although Compton-thick obscuration is associated to outflows in PDS~456, not all Compton-thick obscurations in AGN may necessarily be related to outflows in general. Determining the origin of X-ray obscuration in nearby type-I galaxies provides key observational evidence for understanding the launching mechanisms of outflows in more powerful quasars at higher redshifts, which due to their faint signal cannot be studied with current X-ray observatories. Such winds, with significantly high outflow velocities and mass outflow rates, can play an important role in AGN feedback. The ToO multiwavelength spectroscopy of X-ray eclipses, like performed here on \ngc using \xmm, \nustar, and \hst COS, is an effective way to determine the physical link between the accretion disk, BLR, and outflows in AGN.

\begin{acknowledgements}
This work is based on observations obtained with \xmm, an ESA science mission with instruments and contributions directly funded by ESA Member States and the USA (NASA). This research has made use of data obtained with the \nustar mission, a project led by the California Institute of Technology (Caltech), managed by the Jet Propulsion Laboratory (JPL) and funded by NASA. This work made use of data supplied by the UK \swift Science Data Centre at the University of Leicester. We thank the \swift team for monitoring our AGN sample, and the \xmm, \nustar, and \hst teams for scheduling our ToO triggered observations. SRON is supported financially by NWO, the Netherlands Organization for Scientific Research. This work was supported by NASA through a grant for \hst program number 14481 from the Space Telescope Science Institute, which is operated by the Association of Universities for Research in Astronomy, Incorporated, under NASA contract NAS5-26555. The research at the Technion is supported by the I-CORE program of the Planning and Budgeting Committee (grant number 1937/12). EB acknowledges funding from the European Union's Horizon 2020 research and innovation programme under the Marie Sklodowska-Curie grant agreement no. 655324. SB acknowledges financial support from the Italian Space Agency under grant ASI-INAF I/037/12/0, and from the European Union Seventh Framework Programme (FP7/2007-2013) under grant agreement no. 31278. EC is partially supported by the NWO-Vidi grant number 633.042.525. LDG acknowledges support from the Swiss National Science Foundation. BDM acknowledges support from the European Union's Horizon 2020 research and innovation programme under the Marie Sk{\l}odowska-Curie grant agreement No. 665778 via the Polish National Science Center grant Polonez UMO-2016/21/P/ST9/04025. CP acknowledges support from ERC Advanced Grant Feedback 340442. GP acknowledges support from the Bundesministerium f\"ur Wirtschaft und Technologie/Deutsches Zentrum f\"ur Luft- und Raumfahrt (BMWI/DLR, FKZ 50 OR 1604) and the Max Planck Society. We thank M. Bentz for providing us the host galaxy flux in \ngc, and M. Giustini for useful discussions. We thank the anonymous referee for the useful comments.

\end{acknowledgements}


\newpage
\appendix

\section{Observation logs}
\label{appendix}

%
\begin{table}[!tbph]
\begin{minipage}[t]{\hsize}
\setlength{\extrarowheight}{2pt}
\caption{Log of the triggered \xmm, \nustar, and \hst COS ToO observations of \ngc using our weekly \swift monitoring in 2016.}
\label{obs_table}
\centering
\small
\renewcommand{\footnoterule}{}
\setcounter{LTchunksize}{10}
\setlength{\LTcapwidth}{5.4in}
\begin{tabular}{c c c c c}
\hline \hline
Observatory		&	Obs.	ID			&	Start time	(UTC)	& Length \\
\hline
\xmm			&	0780860901		&	2016-12-11 09:15	& 109.9 ks \\
\xmm			&	0780861001		&	2016-12-21 08:36	& 55.5 ks \\
\nustar			&	80202006002		&	2016-12-11 21:55	& 56.4 ks \\
\nustar 			&	80202006004		&	2016-12-21 10:40	& 45.6 ks \\
\hst COS			& 	LD3E03			&	2016-12-12 11:58	& 2 orbits \\
\hst COS			&	LD3E04			&	2016-12-21 14:23	& 2 orbits \\
\hline
\end{tabular}
\end{minipage}
\end{table}

\end{document}